\documentclass[rnote]{aa} 
\usepackage{graphicx}
\usepackage{txfonts}
\begin{document}
\title{Axions and the pulsation periods of variable white dwarfs revisited}

\author{J. Isern
       \inst{1,2},
       E. Garc\'\i a--Berro
       \inst{2,3} 
       L. G. Althaus 
       \inst{3,4} 
       \and
       A. H. C\'orsico
       \inst{4,5}
       }

\offprints{J. Isern}

\institute{Institut de Ci\`encies de l'Espai (CSIC), 
           Facultat de Ci\`encies, 
           Campus UAB, 
           08193 Bellaterra, 
           Spain\
  	   \and	 
           Institut d'Estudis Espacials de Catalunya, 
           c/ Gran Capit\`{a} 2--4, 
           08034 Barcelona, Spain\         
	   \and	     
	   Departament de F\'\i sica Aplicada, 
           Escola Polit\`ecnica Superior de Castelldefels, 
           Universitat Polit\`ecnica de Catalunya, 
           Avda. del Canal Ol\'\i mpic s/n, 08860 
	   Castelldefels, Spain
           \and
           Facultad de Ciencias Astron\'omicas y Geof\'\i sicas,
           Universidad Nacional de la Plata
           Paseo del Bosque, s/n,
           (1900), La Plata,
           Argentina
           \and
           Instituto de Astrof\'\i sica La Plata, CONICET}
\date{\today}

\abstract{Axions are  the natural  consequence of the  introduction of
          the   Peccei-Quinn   symmetry  to   solve   the  strong   CP
          problem. All  the efforts  to detect such  elusive particles
          have  failed up  to  now.  Nevertheless,  it has  been
          recently shown that the  luminosity function of white dwarfs
          is  best fitted  if axions  with  a mass  of a  few meV  are
          included in the evolutionary calculations.}
         {Our aim is to show  that variable  white dwarfs can  provide additional
          and independent evidence about the existence of axions.}
         {The evolution  of a  white dwarf is  a slow  cooling process
          that  translates into  a secular  increase of  the pulsation
          periods of some variable white dwarfs, the so-called DAV and
          DBV types.  Since axions can freely escape  from such stars,
          their  existence  would   increase  the  cooling  rate  and,
          consequently, the  rate of change of the periods as compared
          with the standard ones.}
         {The present  values of the  rate of change of  the pulsation
          period of  G117--B15A are  compatible with the  existence of
          axions with the masses  suggested by the luminosity function
          of    white    dwarfs,    in    contrast    with    previous
          estimations. Furthermore,  it is  shown that if  such axions
          indeed exist,  the drift of the periods of pulsation  of DBV
          stars would be noticeably perturbed.}
         {}
\keywords{Elementary  particles:  axions  --  stars:  oscillations  --
          stars: white dwarfs}
\authorrunning{Isern et al.}
\titlerunning{Axions and the pulsation periods of variable white dwarfs 
              revisited}
\maketitle


\section{Introduction}

One of the long--standing problems of the  standard model of particles
is the question of why charge and parity can  be violated in 
weak interactions and not
in strong interactions.  This is  the so-called strong CP problem. One
of  the possible  solutions  consists  in the  introduction  of a  new
symmetry, the Peccei-Quinn symmetry  or PQ symmetry, which automatically
accounts  for  the  problem  (Peccei  \&  Quinn  1977a;  1977b).   The
spontaneous breaking of this symmetry gives rise to the existence of a
new particle called  axion, which can couple to  photons, electrons and
baryons  with  a strength  that  depends on  their  mass  and on  the
specific way  in which  the PQ symmetry  is implemented. The  two most
popular implementations are  the KVSZ model (Kim 1979;  Shifman et al.
1980), where the axions couple  with photons and hadrons, and the DFSZ
model (Dine  et al.  1981; Zhitnisky  1980) where they  also couple to
electrons.

From the first  moment that the problem was  formulated, a huge effort
aimed  to directly  detect these  elusive  particles, or  at least  to
constrain  their  properties   using  astrophysical  and  cosmological
arguments, was undertaken. Among  the experiments looking for a direct
detection  we mention  ADMX  and CAST  ---  see Steffen  (2009) for  a
synthetic description.  Excellent reviews about the properties
of  axions  and   the  corresponding  astrophysical  and  cosmological
searches can  be found in Kim  \& Carosi (2008),  Raffelt (2007; 1996;
1990), Khlopov (1999) and Turner (1990).

White  dwarfs represent  the  final evolutionary  stages  of low-  and
intermediate-mass stars.  Since they  are strongly degenerate and
do not  have nuclear sources,  their evolution is just  a gravothermal
process of  cooling.  The simplicity  of these objects, the  fact that
the  physical   processes  necessary  to  understand   them  are  well
identified,  although  not  always  well  known,  and  the  impressive
existing  observational background make  these stars  extremely useful
laboratories to test new ideas  in physics (Isern \& Garc\'\i a--Berro
2008). In particular,  Isern et al.  (2008; 2009) found that the
best fit to  their luminosity function is obtained  when axions of the
appropriate mass  are included in  the calculations.  We note  that in
this case  only DFSZ  axions were taken  into account,  since electron
bremsstrahlung is the dominant process in white dwarf interiors.

  The axion  emission  rate  under these  conditions  is given  by
(Nakagawa et al 1987; 1988):

\begin{equation} 
\dot{\epsilon}_{\rm ax} = 1.08\times 10^{23}\frac{g^2_{\rm aee}}{4\pi}
  \frac{Z^2}{A}T^4_7 F(T,\rho) \,\,\, {\rm erg/g/s},
\end{equation}

\noindent  where $F$ takes  into account  the Coulomb  plasma effects,
$T_7$ is  the temperature in  units of $10^7$  K, $Z$ and $A$  are the
atomic and  mass numbers of  the plasma components,  respectively, and
$g_{\rm aee}$  is the strength of the  axion-electron Yukawa coupling,
which  is not  defined by  the theory  and determines  the  total axion
luminosity of the  white dwarf. The value of  this parameter that best
fits the luminosity function  of white dwarfs is $1.1\times 10^{-13}$,
but  variations  of  a  factor  two  are  still  compatible  with  the
observations (Isern et al. 2008; 2009).

Notice  that  strictly  speaking,   the  shape  of  the  white  dwarf
luminosity function can be  equally well fitted invoking the existence
of any hypothetical light boson  or axion-like particle able to couple
to electrons with a strength $g_{\rm aee} \approx 10^{-13}$. However,
since axions have  a solid theoretical justification, we  will use this
term  all along  the  paper. The  relationship  between this  coupling
constant and the mass of the axion is given by

\begin{equation}
g_{\rm aee} = 2.8 \times 10^{-14} \frac{m_{\rm ax}\cos^2 \beta}{1\, {\rm meV}},
\end{equation}

\noindent where  $\cos^2 \beta$ is an essentially  free parameter that
probably takes a value ranging from  0.5 to 1 and comes from the ratio
of  the two  Higgs fields  that  characterize the  DFSZ axion  models.
Therefore, the axion mass necessary  to fit the white dwarf luminosity
function  mass  is $m_{\rm  ax}\cos^2  \beta  \sim  4$ meV  where,  as
previously mentioned,  variations of $\sim 2$ meV  are compatible with
the  observational  uncertainties.   For  simplicity  we  will  assume
$\cos^2 \beta =1$ and will express our results in terms of the mass
of the axion, as it is usually done. 

Taken at face  value, the result  of Isern  et al. (2009)  can be
considered as  the first indirect proof  that axions of  the DFSZ type
indeed  exist.   However,  before  accepting  this  asseveration  some
conditions  have  to  be   fullfiled:  i)  Different  and  independent
evolutionary codes must provide the  same result, ii) there is no
conventional way  to account for  the presently observed slope  of the
white dwarf  luminosity function, iii) there is  no contradiction
with other  well established astrophysical  or cosmological phenomena,
and  iv) there are  some phenomena  that are  better explained  if the
axion hypothesis  is adopted.   Concerning this last point,  it is
important to realize that the value  of the mass of the axion found in
Isern  et  al.  (2009)  is  close to  the  upper  limit obtained  from
astrophysical constraints.  This means that axions with this mass will
not   dramatically  change  our   current  understanding   of  stellar
evolution, but instead some  subtle changes that accurate observations
could detect  and quantify would be  apparent.  The case  of the white
dwarf  luminosity function  is  just  an example.   The  drift of  the
pulsation periods of  variable white dwarfs, the topic  of this paper,
is another one  that will be discussed in  the following sections.  We
note however that  this is not the case  of core collapse supernovae,
where  axions  with this  mass can  drain  an  important amount  of
gravitational  energy and  modify the  process  of collapse/explosion.
Unfortunately,  the  nucleon-axion interaction,  which  is dominant  in
supernovae, is  not well understood and deserves  additional work that
is  out of the  scope of  this paper.   Finally note  as well  that if
$\cos^2 \beta$ was  small, the mass of the axion  would be higher than
the values quoted here and, consequently, the interaction axion-photon
would be strengthened. This could be used to rule out our results
or to constrain the values that $\cos^2 \beta$ can adopt.

\section{Results and discussion}

During the process of  cooling, white dwarfs experience several phases
of pulsational  instability powered by the  $\kappa$-mechanism and the
``convective driving'' mechanism (Winget \& Kepler 2008). Depending on
their   location  in  the   Hertzsprung--Russell  diagram   and  their
atmospheric  composition they are  called pulsating  PG1159 or  GW Vir
stars, which includes stars with  a surrounding nebula (the PNNV stars)
and stars without nebula (the DOV stars) at $T_{\rm eff} \sim 70,000 -
170,000$ K and  He-, C-, and O-dominated atmospheres,  DBV or V777 Her
stars, with $T_{\rm eff}  \sim 25,000$ K and He-dominated atmospheres,
and  DAV or  ZZ  Ceti stars,  with  $T_{\rm eff}  \sim  12,000$ K  and
H-dominated atmospheres.  All of them display multiperiodic pulsations
with periods  in the range of 100  to 1000 s, although  PNNV stars can
pulsate with periods  as long as 3000 s.  The  length of these periods
clearly  indicates   that  these  objects   are  experiencing  $g$-mode
non-radial pulsations,  where the main restoring force  is gravity ---
see Winget \& Kepler (2008) for a recent review.

   \begin{figure}
   \centering
   \vspace{7cm}
   \includegraphics{13716fig1.eps} 
      \caption{Evolution  of the  measurement of  the period  drift of
               G117--B15A.   Lines  represent  the theoretical  values
               obtained by C\'orsico et  al. (2001) --- solid line ---
               and Bischoff--Kim et  al. (2008) --- dashed-dotted line
               (for the case of a  thin envelope) and dashed line (for
               the   case   of   a   thick  envelope).   The   nominal
               uncertainties of  these models are 1.15,  0.26 and 0.17
               respectively, in the units of the figure.}
      \label{fig}
   \end{figure}

One of  the main characteristics  of $g$-mode pulsations is  that they
experience a  secular drift that can  be used to test  the white dwarf
cooling theory.  This secular drift can be approximately described by

\begin{equation}
\frac{d\ln \Pi}{dt} \simeq -a \frac{d\ln T}{dt} + b \frac{d\ln R}{dt},
\end{equation}

\noindent where $a$  and $b$ are constants of the  order of unity that
depend on  the details of the model,  and $R$ and $T$  are the stellar
radius  and  the  temperature  at  the  region  of  period  formation,
respectively  (Baglin \& Heyvaerts  1969; Winget  et al.  1983).  This
equation  reflects  the  fact  that  as  the  star  cools  down,  the
degeneracy  of  the  plasma  increases, the  buoyancy  decreases,  the
Brunt-V\"ais\"al\"a frequency  becomes lower and,  as a consequence,
the spectrum  of pulsations gradually shifts to  lower frequencies. At
the same time, since the  star contracts, the radius decreases and the
frequency tends to increase. In general, DAV and DBV stars are already
so  cool that  the radial  term is  negligible and  the change  of the
period of pulsation can be directly  related to the change in the core
temperature of the  star. The timescales involved are  of the order of
$\sim 10^{-11}$  s/s for DOVs,  $\sim 10^{-13}$ to $\sim  10^{-14}$ s/
for  DBVs and $\sim  10^{-15}$ to  $\sim 10^{-16}$  s/s for  DAVs. The
measurement of  such drifts is  a difficult task  but, as was 
proved, it is feasible (Winget \& Kepler, 2008).

These  properties  allow us to  build  a  simple  relationship (Isern  et
al. 1992; Isern  \& Garc\'\i a--Berro 2008) to  estimate the influence
of  an extra  sink of  energy,  like axions,  on the  period drift  of
variable white dwarfs:

\begin{equation}
\frac{L_{\rm ax}}{L_{\rm model}} \approx \frac{\dot{\Pi}_{\rm obs}}
{\dot{\Pi}_{\rm model}} -1 \approx 
\frac{\dot{T}_{\rm obs}}{\dot{T}_{\rm model}} -1 ,
\label{eqf}
\end{equation}

\noindent  where the  suffix ``model''  refers to  those  models built
using standard physics.

G117--B15A is a ZZ Ceti star (McGraw \& Robinson 1976) with a dominant
period  of pulsation of  $\sim$ 215  s. This  star, in  an exceptional
observational  effort, has  been observed  for more  than 30  years to
obtain  the secular  evolution  of the  pulsation  period.  Kepler  et
al. (1991a;  1991b), after  $\sim 15$ years  of observations,  found a
secular  drift  of  $\dot{\Pi}_{\rm  obs}  =  (12.0  \pm  3.5)  \times
10^{-15}$ s/s  that was  more than  a factor of  two larger  than the
expected value in the case  of a normal carbon-oxygen white dwarf.  In
order to account  for this discrepancy, Isern et  al.  (1992) proposed
the introduction of  an extra cooling term due to  axions. The mass of
the axions  necessary to fit this  discrepancy was 8.5  meV. Later on,
the analysis of additional ten  years of observations provided a value
of $\dot{\Pi}_{\rm obs}  = (2.3 \pm 1.4) \times  10^{-15}$ s/s (Kepler
et al. 2000), closer to the  value predicted by the standard theory of
evolution  of   white  dwarfs.   A   detailed  seismological  analysis
(C\'orsico et  al. 2001) found  a drift $\dot{\Pi}_{\rm model}  = (3.9
\pm 2.3) \times  10^{-15}$ s/s, where the main  sources of uncertainty
were  due  to the  mode  identification, the  mass  of  the star,  the
chemical profile and the  effective temperature.  This value suggested
that it  was no longer necessary  to invoke axions to  account for the
observed  cooling rate  of white  dwarfs, but  at the  same  time the
uncertainties  were   large  enough   to  prevent  ruling   out  their
existence. It  is important to  realize here that the  values obtained
with Eq.  (\ref{eqf}) agreed well with those obtained from
the accurate numerical calculations of C\'orsico et al. (2001).  A new
analysis  including   additional  five  years   of  observations  gave
$\dot{\Pi} = 3.57 \pm 0.82  \times 10^{-15}$ s/s (Kepler et al. 2005),
a  value  that  overlaps  with  the  theoretical  values  obtained  by
C\'orsico et  al. (2001) and  is marginally compatible with  the axion
luminosity  obtained if  the axion  mass suggested  by  the luminosity
function  of white dwarfs,  $\sim 4$  meV is  adopted.  Finally,  in a
recent release  that includes yet another five years  of observations, Kepler
(2009) obtained  $\dot{\Pi} = 4.77  \pm 0.59  \times 10^{-15}$  s/s.
Adopting the proper motion correction of Kepler et al. (2005), we obtain
 $\dot{\Pi} = 4.07  \pm 0.61  \times 10^{-15}$  s/s, a value that is consistent with the  existence of the extra cooling term suggested by the
luminosity function.  Figure  \ref{fig} displays the evolution of
the  measured value  of $\dot\Pi$  with time.   The jumps  between the
different   observational   values   reflect   the   introduction   of
corrections, like the proper motion  correction, and the fact that the
phase  of  the   periodicity  is  affected  by  a   jitter  caused  by
low-amplitude  modes, contamination  by G117--B15B, the companion star, 
 use  of different
telescopes  and  apertures\ldots  that  are not  well  understood  yet
(Kepler, private communication).

\begin{table}
\caption{Axion luminosities and axion  masses necessary to account for
         the observed drift of the period of pulsation of G117--B15A.}
\label{tabg}
\centering
\begin{tabular}{lcc}
\hline
\hline
Model & $L_{\rm ax}/ 10^{30}$ (erg/s) & $m_{\rm ax} \cos^2\beta$ (meV) \\
\hline
C           & $0.6 (2.8/0.0)$ &  2.8   (6.0/0.0) \\
BK (thick)  & $6.8 (8.7/4.9)$ &  9.3 (10.6/7.9) \\
BK (thin)   & $2.2 (3.5/1.0)$ &  5.5 ( 6.9/3.7) \\
\hline
\hline
\end{tabular}
\end{table}

Table \ref{tabg}  displays the axion luminosity and  mass necessary to
fit the  observations when several white dwarf  pulsational models are
adopted  (the values  in  brackets correspond  to the  observational
uncertainties).  The  values of the  first row, labeled model  C, were
obtained using Eq. (\ref{eqf}), the model of G117--B15A that best fits
the data (C\'orsico  et al.  2001) and the  previously mentioned axion
emission rates  of Nakagawa  et al.  (1987;  1988). It is important  
to realize here that  in a completely
independent  analysis,  Bischoff-Kim  et  al.  (2008)  identified  two
possible  asteroseismological models  of G117--B15A,  one with  a thin
hydrogen  envelope and  the  other with  a  relatively thick  hydrogen
envelope, with  $\dot{\Pi} =  2.98 \pm 0.17  \times 10^{-15}$  s/s and
$1.92\pm  0.26 \times  10^{-15}$  s/s, respectively.  Both values  are
smaller than the measured ones and indicate that an additional sink of
energy  is necessary. Table  \ref{tabg} shows  the luminosity  and the
mass of  the axions necessary to  account for the most  recent data of
Kepler  (2009)   assuming  a  temperature  of  the   core  of  $T_{\rm
c}=1.2\times 10^7$ K in both cases.

Recently, Mukadam et  al. (2009) analyzed R  548 (ZZ Ceti itself)
and found  that its mass lies between  0.55 and $0.58\, M_{\sun}$,
while the mass of the hydrogen envelope is $\log M_{\rm H}\simeq -5.0$ and
that  the best  prediction for  the rate  of change  of the  period is
$\dot{\Pi} =(8.3  \pm 0.046) \times  10^{-15}$ s/s. These  results are
fully consistent with the values  obtained for G117--B15A and with those
 displayed in Table \ref{tabg}.

The pulsation properties  of DBV stars can provide  an additional test
to the  axion hypothesis.  Since these  stars have a  hotter core than
that  of DA  white dwarfs,  the neutrino  luminosity can no  longer be
neglected. and the influence of axions on the pulsation period drift is
smaller than  that expected  for DA white  dwarfs. For instance,  a DB
white dwarf  of $  M \sim 0.59\,  M_{\sun}$, an  effective temperature
$T_{\rm eff}\sim 25\,200$  K and a core temperature  $\log T_{\rm c} =
7.61$ has a luminosity of $L \sim 6.3 \times 10^{-2}\, L_{\sun}$ and a
neutrino luminosity $L_\nu \sim  5.1 \times 10^{-2}\,L_{\sun}$.  If an
axion mass of  5 meV is assumed, the  axion luminosity amounts $L_{\rm
  ax} =  6.0 \times 10^{-2}\, L_{\sun}$ and  the corresponding secular
change of the pulsation period would be

\begin{equation}
\frac{\dot{\Pi}_{\rm ax}}{\dot{\Pi}_{\rm no-ax}}\simeq 1.5 \, .
\end{equation}  

\noindent Note that because of the strong dependence of the different
luminosities  on  the  core   temperature,  this  prediction  is  very
sensitive  to  the  position  in  the instability  strip  and  on  the
particular parameters  of the  star under study.  The only case  of DB
variable white dwarf that  is currently being studied is EC20058--5234
(Sullivan 2009), but unfortunately a  rate of period change is not yet
available for this star.

\section{Conclusions}

It has been shown that white dwarf variable G117--B15As present value 
of the secular rate of change of the  period of pulsation, 
 $4.07  \pm 0.61 \times  10^{-15}$ s/s (Kepler 2009), is consistent with
the predictions  of the theoretical  models, as was  anticipated by
Isern et al.   (1992), if an additional source  of cooling like axion
emission is included.  This  result is corroborated by the completely
independent  analysis of  the same  star done  by Bischoff-Kim  et al.
(2008), using the measured rate of period change ($3.57 \pm 0.82 \times
10^{-15}$ s/s)  obtained previously (Kepler et al.  2005).  This means
that  the  conclusion of  C\'orsico  et al.  (2001)  that  it was 
unnecessary to introduce an additional cooling source is no longer valid
when the  new observational  data are taken  into account.  This  is an
important  result, as  asteroseismological observations of white dwarfs
  seem  to give
additional and independent support to the claim of Isern et al. (2008;
2009) that the white dwarf luminosity  function is better fitted if an
additional cooling  provided by  axions with  a mass of  a few  meV is
included in the calculations.

It has also  been shown that since the core  temperatures of DBV stars
are higher, the secular drift of the pulsation period of these stars is
strongly  enhanced by  the axion  emission.  In  particular,  if axion
emmisivity is  computed using the mass  obtained from the  best fit to
the  luminosity   function  of  white  dwarfs  and   included  in  the
calculations,  the secular  rate of  period change  increases  by $\sim
50\%$, although the  exact value depends on the  core temperature and,
consequently, on their position in the DB variability strip.

Finally,  it is worthwhile  to emphasize  the importance  of continued
obervations of  G117--B15A and  R 548. It is also  important to
enlarge  the number  of  stars  with observed  period  drifts and  the
measurement of these drifts in hotter variables like DBV white dwarfs.
Finally, it  is also  important to improve  the asteroseismological
models.  All  these actions would  undoubtely contribute to  solve the
problem of axions.

\begin{acknowledgements}
This  work  was  supported  by  the MCINN  grants  AYA08-1839/ESP  and
AYA2008-04211-C02-01 and the AGAUR grants SGR1002/2009 and SGR315/2009
of the Generalitat de Catalunya and by the European Union FEDER funds.
LGA also acknowledges a PIV grant of the AGAUR  of the  Generalitat de
Catalunya.
\end{acknowledgements}

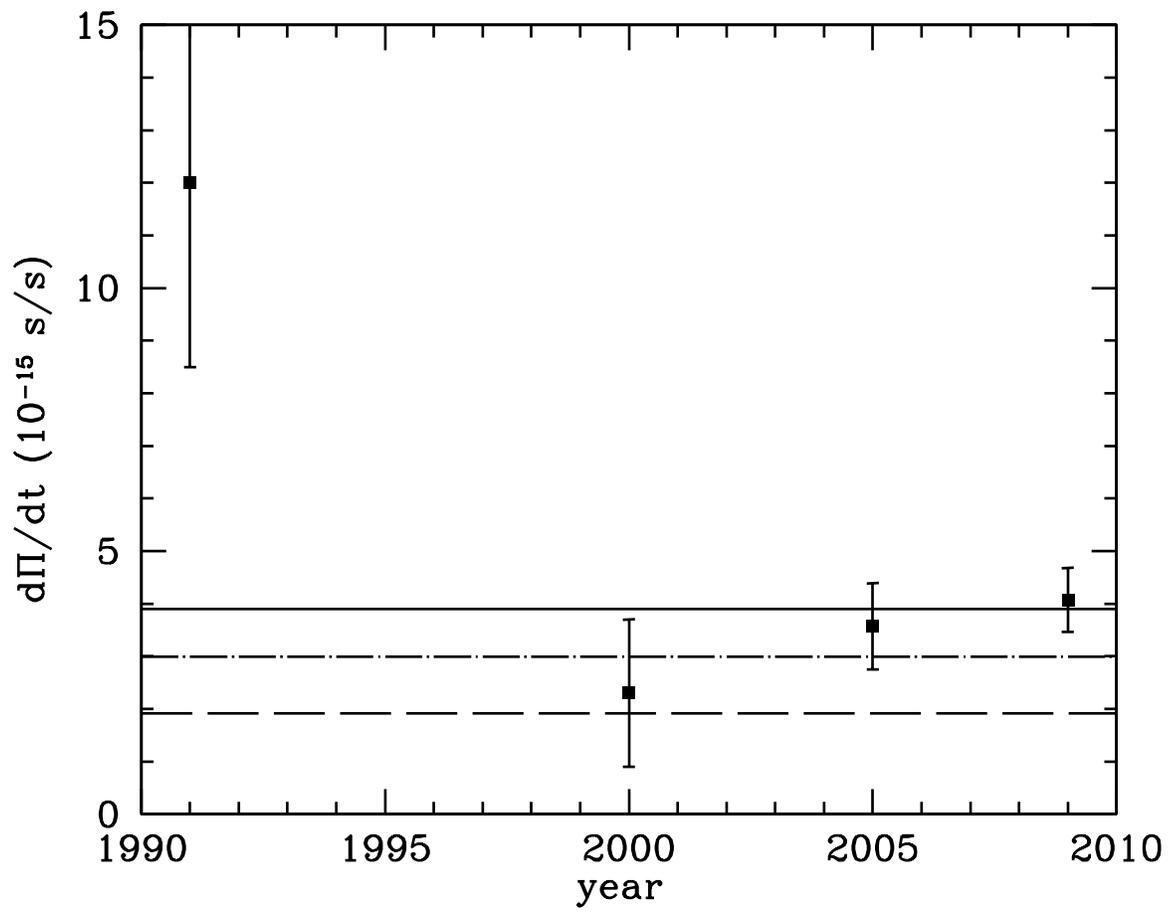
\end{document}